\documentclass[a4paper]{JHEP3}
\usepackage{latexsym}
\usepackage{amsfonts}
\usepackage{epsfig}

\title{Poincar\'e Dual of D=4 N=2 Supergravity with Tensor Multiplets} 

\author{Luca Sommovigo\\ Dipartimento di Fisica,
  Politecnico di Torino,\\ Corso Duca degli Abruzzi 24, I-10129
  Torino, and\\ Istituto Nazionale di Fisica Nucleare (INFN)\\ Sezione
  di Torino, Italy\\ E-mail: \email{luca.sommovigo@polito.it}} 

\abstract{We study, in an arbitrary even number $D$ of dimensions, the
  duality between massive $D/2$ tensors coupled to vectors, with
  masses given by an arbitrary number of ``electric'' and ``magnetic''
  charges, and $(D/2-1)$ massive tensors. We develop a formalism to
  dualize the Lagrangian of $D=4$, $N=2$ supergravity coupled to
  tensor and vector multiplets, and show that, after the dualization,
  it is equivalent to a standard $D=4$, $N=2$ gauged supergravity in
  which the Special Geometry quantities have been acted on by a
  suitable symplectic rotation.}  
 
\begin{document}

\bibliographystyle{luca}

\def\op#1{\mathcal{#1}}

\def\IC{\relax\,\hbox{$\inbar\kern-.3em{\rm C}$}}
\def\inbar{\vrule height1.5ex width.4pt depth0pt}
\def\bfone{\relax{\rm 1\kern-.35em 1}}
\def\bfnull{\relax{\rm O \kern-.635em 0}}

\def\A{\mathcal{A}}
\def\a{\alpha}
\def\b{\beta}
\def\bos{{\rm bos}}
\def\c{\chi}
\def\cb{\bar{\chi}}
\def\cc{{\rm c.c.}}
\def\d{\delta}
\def\D{\Delta}
\def\de{{\rm d}}
\def\der{\partial}
\def\dx{\right}
\def\e{\epsilon}
\def\ee{\left( e^2 \right)}
\def\eet{\left( {\tilde e}^2 \right)}
\def\esm{\left( \frac{e}{m} \right)}
\def\etsm{\left( \frac{{\tilde e}}{m} \right)}
\def\et{{\tilde e}}
\def\F{\mathcal{F}}
\def\Fh{\hat\mathcal{F}}
\def\g{\gamma}
\def\G{\Gamma}
\def\H{\mathcal{H}}
\def\i{{\rm i}}
\def\ib{{\bar\imath}}
\def\im{{\rm Im}\,\mathcal{N}}
\def\imez{\frac{{\rm i}}{2}}
\def\jb{{\bar\jmath}}
\def\kb{{\bar k}}
\def\l{\lambda}
\def\L{\Lambda}
\def\La{\mathcal{L}}
\def\lb{{\bar \ell}}
\def\m{\mu}
\def\M{\mathcal{M}}
\def\mb{{\bar m}}
\def\me{\left( em \right)}
\def\met{\left( {\tilde e}m \right)}
\def\mez{\frac{1}{2}}
\def\mm{\left( m^2 \right)}
\def\mq{\left( e^2 + m^2 \right)}
\def\mqt{\left( {\tilde e}^2 + m^2 \right)}
\def\n{\nu}
\def\N{\mathcal{N}}
\def\Nb{\mathcal{\overline{N}}}
\def\na{\nabla}
\def\o{\omega}
\def\O{\Omega}
\def\ol{\overline}
\def\qu{\frac{1}{4}}
\def\R{\mathcal{R}}
\def\r{\rho}
\def\rb{{\bar r}}
\def\re{{\rm Re}\,\mathcal{N}}
\def\s{\sigma}
\def\S{\Sigma}
\def\sb{{\bar s}}
\def\sx{\left}
\def\t{\tau}
\def\th{\theta}
\def\Th{\Theta}
\def\U{\mathcal{U}}
\def\ve{\varepsilon}
\def\w{\wedge}
\def\z{\zeta}

\newcommand{\nn}{\nonumber}
\newcommand{\noi}{\noindent}


\section{Introduction}
\label{sec:intro}

It is a well known fact that orientifold compactifications of IIB
string theory with fluxes turned on, give rise to a theory in which
tensor multiplets are present, and are endowed with a mass; this fact
has recently given rise to a renewed interest in theories describing
tensor multiplets coupled to scalar and vector multiplets
\cite{Louis:2004xi,Kuzenko:2004tn,Theis:2004pa}.\\ 
This same topic has been considered also in \cite{D'Auria:2004sy}
where a theory describing, in $D$ even dimensions, a number of massive
$D/2$--form tensors $B_I$ with masses given by ``magnetic'' and
``electric'' couplings has been dualized to a theory of massive
$D/2-1$--dimensional form fields, generalizing the well-known duality
in 4 dimensions between massive 2--form tensors and massive vectors.  
The ``electric'' and ``magnetic'' couplings which give mass to the
tensors arise naturally from Fayet-Iliopoulos terms in the context of
the dualization of $n_T$ massless scalars to 2--forms in $D=4$, $N=2$
supergravity \cite{Dall'Agata:2003yr,D'Auria:2004yi,Sommovigo:2004vj}.
In this light, the theory of \cite{D'Auria:2004sy} can be thought of
as a generalization to an arbitrary number of dimensions of the
linearized (without coupling to scalars) bosonic kinetic and mass
terms for the $n_T$ 2--forms $B_I$ from $D=4$, $N=2$ supergravity
coupled to $n_T$ tensor multiplets and $n_V$ vector multiplets
\cite{D'Auria:2004yi}.\\
Actually in \cite{D'Auria:2004sy} the proof of this generalized
duality has been given only for two particular cases: that of a single
tensor $B$, when $D=4d$ and hence $B$ is a $2d$--form field with
coupling constants $m$ and $e$, and that of a couple of tensors, $B_I$
with $I=1,2$ when $D=4d+2$ and hence the $B_I$'s are $(2d+1)$--form
fields, taking the coupling constants to be $m^{I\L} = m \d^{I\L}$ and
$e^I_\L = e \e^I_\L$, with $\L=1,2$.\\ 
In both cases, the squared mass of the $(D/2-1)$--form after the
dualization has been shown to depend on both ``electric'' and
``magnetic'' charges, via the relation:

\begin{equation}
  \label{mu}
  \m = \sqrt{e^2 + m^2},
\end{equation}

\noi so that no distinction appears between electric and magnetic
charges. We will extend the analysis of \cite{D'Auria:2004sy} to the 
more general case in which an arbitrary number of ``electric'' and
``magnetic'' charges is present, in the linearized theory, showing
that also in this case the Poincar\'e dual of a theory of $n_T$
massive $D/2$--form fields is  a theory describing $n_T$ massive
$D/2-1$--form fields. Moreover their square mass matrix will be given
by the ``sum of the squares'' of the electric and magnetic charge
matrices: 

\begin{equation}
  (\m^2)^{IJ} = \sx( e^2 + m^2 \dx)^{IJ} \equiv \sx( e^I_\L \d^{\L\S}
  e^J_\S + m^{I\L} \d_{\L\S} m^{J\S} \dx) \label{massq}
\end{equation}

\noi thus generalizing relation (\ref{mu}).\\ 
The whole procedure requires the matrices $e^I_\L$, $m^{I\L}$ and
$\mq^{IJ}$ to be invertible, therefore from now on we will suppose
that $n_T = n_V = n$, so that all these matrices are square, and that
$\det e$, $\det m$, $\det \mq \neq 0$ in order to guarantee the
existence of the inverses.\\
The paper is organized as follows: in section \ref{sec:dual} we
prepare the setting and write down the dual Lagrangian for a theory of
$n$ massive tensors in an arbitrary even number $D$ of dimensions,
distinguishing between $D=4d$ and $D=4d+2$ cases, generalizing the
procedure of \cite{D'Auria:2004sy}.\\
In section \ref{sec:stsugra} we apply the whole formalism to the
Lagrangian of $D=4$ $N=2$ supergravity coupled to $n$ massive tensors
and $n$ vectors obtained in \cite{D'Auria:2004yi}.\\
Finally, in section \ref{syrot} we reinterpret the dual Lagrangian
obtained with the dualization as a standard $D=4$ $N=2$ gauged
supergravity, where the gauge is purely electric and the quantities of
the special geometry can be obtained from the standard ones by means
of a symplectic rotation.\\
We note that the fact that the $N=2$ tensor coupled theory can be
related to a standard $N=2$ supergravity by means of duality
transformations does not mean that they are the same theory. As a
matter of fact we can show that they give rise to different $N=1$
theory when a suitable truncation to $N=1$ is performed (see reference
\cite{wip}).

\section{General dualization procedure for the linearized theory}
\label{sec:dual}

In this section we generalize the procedure of reference
\cite{D'Auria:2004sy} to obtain the dual Lagrangian of a theory of
massive tensors in an arbitrary even number of dimensions with an
arbitrary number of charges.\\
\noi The linearized Lagrangian of a number $n$ of massless vector 
fields and a number $n$ of tensor fields, as coming from $D=4$ $N=2$
supergravity \cite{Dall'Agata:2003yr,D'Auria:2004yi}, is\footnote{Note
  that, in order to make contact with \cite{D'Auria:2004yi} one has to
  redefine the ``electric'' and ``magnetic'' charges as $2 \, e^I_\L
  \to e^I_\L$, $2 \, m^{I\L} \to m^{I\L}$.}:

\begin{eqnarray}
  \La^{T,\, V} &=& - \qu \d^{IJ} \H_I \w {}^* \H_J - e^I_\L \sx( 
  \F^\L + \mez m^{J\L} B_J \dx) \w B_I \nn \\
  && + \mez \d_{\L\S} \sx( \F^\L + m^{I\L} B_I \dx) \w {}^* \sx(
  \F^\S + m^{I\S} B_J \dx) \label{schla}
\end{eqnarray}

\noi where $\d$ is the usual Kronecker delta, $\F^\L$, $\H_I$ are the
field-strengths of the gauge vectors $A^\L$ and the tensors $B_I$
respectively.\\  
Lagrangian (\ref{schla}) can be immediately generalized to an
arbitrary (even) number $D$ of dimensions by promoting $B_I$ and
$\F^\L$ to be $D/2$--form fields\footnote{We adopt a mostly minus
  metric $(+,-,\dots,-)$ so that, in order to have positive kinetic
  energy, when $D=4d+2$ an overall minus sign in front of the
  Lagrangian is needed.}. As is well known, Lagrangian (\ref{schla})
and its generalization to higher dimensions are invariant under the
following gauge transformations:

\begin{equation}
  \label{gaugtr}
  \d A^\L = -m^{I\L} \O_I, \qquad \d B_I = \de \O_I,
\end{equation}

\noi where $\O_I$ is a $D/2-1$-form field, if and only if the
generalized tadpole cancellation condition:

\begin{equation}
  \label{tcc}
  e^I_\L m^{J\L} \mp e^J_\L m^{I\L} = 0
\end{equation}

\noi holds. The minus sign holds in the $D=4d$ case, when the $B_I$ are
even forms (and hence they commute with each other), while the plus
sign holds in the $D=4d+2$ case, when the $B_I$ are odd forms
(anticommuting with each other).\\
It will prove convenient to add to (\ref{schla}) a total derivative:   

\begin{equation}
  \label{td}
  - \mez e^I_\L \sx( m^{-1} \dx)_{I\S} \F^\L \w \F^\S,
\end{equation}

\noi in such a way that the topological term proportional to the
``electric'' charge $e^I_\L$ can be written in a manifestly gauge
invariant way:

\begin{equation}
  \label{mgi}
  - \mez e^I_\L \sx( m^{-1}\dx)_{I\S} \sx( \F^\L + m^{J\L} B_J \dx)
    \w \sx( \F^\S + m^{K\S} B_K \dx).
\end{equation}

\noi By means of a gauge transformation, the field-strengths of the
vectors can be reabsorbed into the tensors: 

\begin{equation}
  \O_I = \sx( m^{-1} \dx)_{I\L} A^\L \, \Rightarrow \, {A'}^\L \equiv
  A^\L + \d A^\L = 0; \qquad B_I' \equiv B_I + \sx( m^{-1} \dx)_{I\L}
  \F^\L \label{redg}
\end{equation}

\noi and the Lagrangian can be written in the simpler form:

\begin{equation}
  \label{lagsim}
  \pm \La^{T,\, V} = - \qu \d^{IJ} \H_I \w {}^* \H_J + \mez \mm^{IJ}
  B_I \w {}^* B_J - \mez \me^{IJ} B_J \w B_I    
\end{equation}

\noi the two signs referring respectively to the case $D=4d$ and
$D=4d+2$. We have defined the following matrices:

\begin{eqnarray}
  \mm^{IJ} &\equiv& m^{I\L} \d_{\L\S} m^{J\S} \label{defmm} \\
  \me^{IJ} &\equiv& e^I_\L m^{J\L}. \label{defem}
\end{eqnarray}

\noi It should be noted that both matrices have a well-defined
symmetry in the indices $I,J$: $\mm^{IJ}$ is symmetric by construction
while $\me^{IJ}$, due to the tadpole cancellation condition
(\ref{tcc}) is symmetric if $D=4d$ and antisymmetric if $D=4d+2$.\\  
Starting from the Lagrangian of equation (\ref{lagsim}), the process
of dualization requires that $\H_I$ and $B_I$ be considered as
independent fields, enforcing the relation $\H_I = \de B_I$ by means
of the equations of motion for a suitable Lagrange multiplier $\r^I$
that should be added to the Lagrangian, which becomes:

\begin{eqnarray}
  \pm \La^{T,\, V} &=& - \qu \d^{IJ} \H_I \w {}^* \H_J + \mez \mm^{IJ}
  B_I \w {}^* B_J - \mez \me^{IJ} B_J \w B_I + \nn \\  
  && + \r^I \w \sx( \H_I - \de B_I \dx). \label{lagmul}
\end{eqnarray}

\noi In order to get the dualized Lagrangian, at first one has to
write down the equations of motion for the independent fields $B_I$
and $\H_I$, then to solve these equations in terms of the dual field
$\r^I$ and finally to substitute the resulting expressions into the
original Lagrangian.\\
Because of the (anti)-symmetry properties of the matrix $\me^{IJ}$ and
the fact that the $B_I$ and $\H_I$ fields are even or odd forms
depending on the fact that $D=4d$ or $D=4d+2$, it is more
convenient to consider the two cases separately. 

\subsection{$D=4d$}
\label{subsd4d}

When $D=4d$, the $B_I$'s are $2d$--forms, and hence they commute with
all other $p$--form fields; the equations of motion are:

\begin{eqnarray}
  \mez \d^{IJ} {}^* \H_J + \r^I &=& 0 \label{eomh}\\
  \mm^{IJ} {}^* B_J - \me^{IJ} B_J + \de \r^I &=& 0 \label{eomb} 
\end{eqnarray}

\noi Equation (\ref{eomh}) is easily invertible in order to get $\H_I$
in terms of ${}^* \r^I$, while from equation (\ref{eomb}) and its
Hodge dual it is possible to write $B_I$ in terms of both $\de \r^I$
and ${}^* \de \r^I$ as: 

\begin{eqnarray}
  B_I &=& \mq^{-1}_{IJ} \sx( {}^* \de \r^J + \esm^J_K \de \r^K \dx)
  \label{brho} \\
  {}^* B_I &=& - \mq^{-1}_{IJ} \sx( \de \r^J - \esm^J_K {}^* \de \r^K
  \dx) \label{bsrho}
\end{eqnarray}

\noi where the following matrices have been introduced:

\begin{eqnarray}
  \esm^I_J &\equiv& e^I_\L \d^{\L\S} \sx( m^{-1} \dx)_{J\S}
  \label{esmdef} \\
  \ee^{IJ} &\equiv& e^I_\L \d^{\L\S} e^J_\S = \ee^{JI}. \label{eedef}  
\end{eqnarray}

\noi As $\mm^{IJ}$, also $\ee^{IJ}$ is symmetric by construction. Using these
symmetries, together with the following identities involving the
product of two matrices: 

\begin{eqnarray}
  \me^{IK} \esm^J_K &=& \ee^{IJ} \label{emesm}\\
  \esm^I_K \ee^{KJ} &=& \esm^I_K \me^{KL} \esm^J_L = \esm^J_K \ee^{KI}
  \label{esmee} 
\end{eqnarray}

\noi which allow a number of simplifications, the dualized
Lagrangian turns out to be:

\begin{equation}
  \La^{dual} = \mez \mq^{-1}_{IJ} \de \r^I \w {}^* \de \r^J + \mez
  \mq^{-1}_{IK} \esm^K_J \de \r^I \w \de \r^J - \d_{IJ} \, \r^I \w
  {}^* \r^J. \label{lagdual}
\end{equation}

\noi As expected, (\ref{lagdual}) is the Lagrangian describing a
theory of $n$ massive $(2d-1)$ vector fields, whose mass matrix is
$(\m^2)^{IJ} = \mq^{IJ}$\footnote{Here and in the next case, the
  topological term has been written for completeness, even if its
  presence does not modify the equations of motion.}.

\subsection{$D=4d+2$}
\label{subsd4d+2}

In this case, the $B_I$ fields are $2d+1$--forms, hence anticommuting
with all other odd forms; taking into account these different
properties, we can obtain the equations of motion:

\begin{eqnarray}
  - \mez \d^{IJ} {}^* \H_J + \r^I &=& 0 \label{eomhII} \\
  \mm^{IJ} {}^* B_J - \me^{IJ} B_J - \de \r^I &=& 0 \label{eombII}
\end{eqnarray}

\noi again, as in the previous case, the equations of motion for the
$\H_I$ fields are easily inverted, while after some algebraic
manipulation on the equations for the $B_I$ and their Hodge dual,
remembering that $\me^{IJ}$ is now an antisymmetric matrix, and hence

\begin{equation}
  \esm^I_K \ee^{KJ} = \esm^I_K \me^{KL} \esm^J_L = - \esm^J_K \ee^{KI}
  \label{esmee2} 
\end{equation}

\noi we get the relations: 

\begin{eqnarray}
  B_I &=& \mq^{-1}_{IJ} \sx( {}^* \de \r^J + \esm^J_K \de \r^K \dx)
  \label{brhoII} \\
  {}^* B_I &=& \mq^{-1}_{IJ} \sx( \de \r^J + \esm^J_K {}^* \de \r^K
  \dx) \label{bsrhoII}
\end{eqnarray}

\noi Using (\ref{brhoII}) and (\ref{bsrhoII}) in (\ref{lagmul}), we
get the following expression: 

\begin{equation}
  \La^{dual} = - \mez \mq^{-1}_{IJ} \de \r^I \w {}^* \de \r^J - \mez
  \mq^{-1}_{IK} \esm^K_J \de \r^I \w \de \r^J + \d_{IJ} \r^I \w {}^*
  \r^J. \label{lagdualII}
\end{equation}

\noi which is the Lagrangian describing $n$ massive vectors in
$D=4d+2$ dimensions. We have therefore shown that, in an even number
$D$ of dimensions, a theory of $n$ massive $D/2$--form tensors is dual 
to a theory of $n$ massive $D/2 -1$--form tensors.

\section{The $D=4$ $N=2$ Supergravity theory coupled to vector
  multiplets and scalar--tensor multiplets} 
\label{sec:stsugra}

The procedure seen in section \ref{sec:dual} can be applied to the
Lagrangian of the $D=4$ $N=2$ supergravity theory coupled to $n_V$
vector multiplets and $n_T$ scalar-tensor multiplets
\cite{D'Auria:2004yi}. The first step is to write the Lagrangian of
reference \cite{D'Auria:2004yi} in first-order formalism, which is
more practical to perform the dualization. Actually we do not need
all the Lagrangian, but only the terms coupled either to $\H_I$ or
$B_I$, which are collected in $\La_\bos$ and $\La_{\rm Pauli}$:  

\begin{equation}
  \La^{T,\, V}_{D=4} = \La_\bos + \La_{\rm Pauli} \label{lag}
\end{equation}

\begin{eqnarray}
  \label{lagbos}
  \La_\bos &=& - \mez \im_{\L\S} \, \Fh^\L \w {}^* \Fh^\S + \mez
  \re_{\L\S} \, \Fh^\L \w \Fh^\S - \qu \M^{IJ} \H_I \w {}^* \H_J +
  \nn \\ 
  && + \H_I \w A^I - e^I_\L \sx( \Fh^\L - \mez m^{J\L} B_J \dx) \w B_I
  \\
  \La_{\rm Pauli} &=& - \imez \, \H_I \w {}^* \sx( \U^{IA\a} \ol
  \psi_A \w \g \w \g \, \z_\a - \U^I{}_{A\a} \ol \psi^A \w \g \w \g \,
  \z^\a \dx) + \nn\\ 
  && - \i \, \H_I \w \D^{I\a}{}_\b \, \ol \z_\a \g \, \z^\b - \H_I \w 
  \sx( \U^{IA\a} \ol \psi_A \, \z_\a + \U_{J\,A\a} \ol \psi^A \, \z^\a
  \dx) + \nn \\ 
  && - 2 \, \i \, \im_{\L\S} \, \Fh^\L \w \sx[ L^\S \sx( \ol \psi^A \w
  \psi^B \e_{AB} \dx)^- - \ol L^\S \sx( \ol \psi_A \w \psi_B \e^{AB}
  \dx)^+  \dx] + \nn \\    
  && - 2 \, \im_{\L\S} \, \Fh^\L \w \sx[ \ol f_\ib^\S \sx( \ol
  \l^\ib_A \g \w \psi_B \e^{AB} \dx)^- + f_i^\S \sx( \ol \l^{iA} \g \w
  \psi^B \e_{AB} \dx)^+ \dx] + \nn \\     
  && - \frac{\i}{4} \, \im_{\L\S} \, \Fh^\L \w \sx( \na_i f_j^\S \ol
  \l^{iA} \g \w \g \, \l^{jB} \e_{AB} - \na_\ib \ol f_\jb^\S \ol
  \l^\ib_A \g \w \g \, \l^\jb_B \e^{AB} \dx) + \nn \\
  && + \imez \, \im_{\L\S} \, \Fh^\L \w \sx( L^\S \, \ol \z_\a \g \w
  \g \, \z_\b \, \IC^{\a\b} - \ol L^\S \, \ol \z^\a \g \w \g \, \z^\b
  \, \IC_{\a\b} \dx) \label{lagpauli}
\end{eqnarray}

\noi where we have defined the following quantities:

\begin{eqnarray}
  \Fh^\L &\equiv& \F^\L + m^{I\L} B_I \label{deffhat} \\
  A^I &\equiv& A^I_u \de q^u \label{offdmet} \\
  \g &\equiv& \g_\m \de x^\m \label{gamma} \\
  \sx( \dots \dx)^\pm &\equiv& \sx( \dots \dx) \pm \i {}^* \sx( \dots 
  \dx) \label{pm}  
\end{eqnarray}

\noi And the bilinear fermions $\sx( \dots \dx)^\pm$ enjoy the
following property: 

\begin{equation}
  {}^* \sx( \dots \dx)^\pm = \mp \i \sx( \dots \dx)^\pm \label{prop} 
\end{equation}

\noi the Hodge dual operator ${}^*$ being defined as:

\begin{equation}
  {}^* \de x^{\m_1} \wedge \dots \wedge \de x^{\m_k} \equiv
  \frac{1}{(n-k)!} \sqrt{-g} \e^{\m_1 \dots \m_k}{}_{\m_{k+1} \dots 
  \m_n} \de x^{\m_{k+1}} \wedge \dots \wedge \de
  x^{\m_n}. \label{hodge}
\end{equation}

\noi Moreover $\M_{IJ}$ and $A^I_u$ are the remnants of the
quaternionic metric $h_{\hat u \hat v}$ according to the decomposition  
given in \cite{Theis:2003jj,Dall'Agata:2003yr}:

\begin{equation}
  h_{{\hat u}{\hat v}} = \begin{pmatrix}{g_{uv} + \M_{IJ} A^I_u A^J_v 
  & \M_{IK} A^K_u \cr \M_{IK} A^K_v & \M_{IJ}}\end{pmatrix}.
  \label{quamet} 
\end{equation}

\noi $L^\L$ and $\ol f^\L_\ib$ are the upper components of the
symplectic sections of standard $N=2$ supergravity, while $\im_{\L\S}$
and $\re_{\L\S}$ are the imaginary and real parts of the period matrix
$\N_{\L\S}$ of the special geometry; $\U_{IA\a}$ and $\D_I{}^\a{}_\b$
are the remnants of the quaternionic vielbein and symplectic
connection in the dualized directions. ($\z^\a$) $\z_\a$ are the
(anti)-chiral hyperinos, ($\l^\ib_A$) $\l^{iA}$ are the (anti)-chiral
gauginos and ($\psi^A$) $\psi_A$ are the (anti)-chiral
gravitinos. Finally, $\e^{AB}$ is the totally antisymmetric tensor and
$\IC^{\a\b}$ is the charge conjugation matrix.\\ 
In order for the theory to be supersymmetric, as a consequence of the
Ward identity of supersymmetry, the electric and magnetic charges must
satisfy the constraint 

\begin{equation}
  e^I_\L m^{J\L} - e^J_\L m^{I\L} = 0, \label{tcc4d}
\end{equation}

\noi which coincides with equation (\ref{tcc}) in the case $D=4d$. 
Lagrangian (\ref{lag}) is invariant under the gauge transformations
parametrized by a 1--form field $\O_I$ already described in
(\ref{gaugtr}). As explained in the previous section, we can add to
the Lagrangian a topological term which does not modify the equations
of motion but provides a more compact expression for the terms of the
type $\F \w \F$. \\ 
The invariance of the action is now explicit, due to the fact that
the fields $B_I$ and $\F^\L$ appear only through either $\Fh^\L$ or
$\H_I$, which are invariant under (\ref{gaugtr}) by construction,
since: 

\begin{eqnarray}
  \d \Fh^\L &=& \de \d A^\L + m^{I\L} \d B_I = - m^{I\L} \O_I +
  m^{I\L} \O_I = 0 \label{delf} \\ 
  \d \H_I &=& \de \d B_I = \de \de \O_I =0 \label{delh} 
\end{eqnarray}

\noi The effect of the gauge fixing of equation (\ref{redg}) is the
replacement of $\Fh^\L$ with $m^{I\L} B_I$.\\
Collecting all the fermionic bilinears, it is possible to rewrite the
Lagrangian in a more compact expression:  

\begin{eqnarray}
  \label{lagIV}
  \La^{T,\, V}_{D=4} &=& - \qu \M^{IJ} \H_I \w {}^* \H_J + \H_I \w A^I
  + \H_I \w S^I - \mez \mm^{IJ} B_I \w {}^* B_J + \nn \\
  && - \mez \met^{IJ} B_I \w B_J + B_I \w T^I + \r^I \w \sx( \H_I - \de
  B_I \dx)
\end{eqnarray}

\noi where $\mm^{IJ}$ is defined as in (\ref{defmm}) with the matrix
$\im$ instead of $\d$. Notice that $\im$ is negative definite, and
hence the sign of the term proportional to $\mm^{IJ}$ is different from
that in (\ref{lagmul}). We have also introduced the following quantity:

\begin{equation}
  \et^I_\L = e^I_\L - \re_{\L\S} m^{I\S} \label{etdef}
\end{equation}

\noi and $\met^{IJ}$ is defined as in (\ref{defem}) replacing $e^I_\L$
with $\et^I_\L$. The $T^I$ are the Pauli terms arising from the
coupling of the fermions to $\Fh$, while $S^I$ is the remnant of the
quaternionic Pauli terms coupled to the differential $\de q^I$ of the
axionic coordinates which have been dualized. Notice that $\met^{IJ}$
is symmetric in the indices $I,J$ because of the tadpole cancellation
condition and of the symmetry of the real part of the period matrix
$\N_{\L\S}$.\\ 
\noi From Lagrangian (\ref{lagIV}) it is easy to write down the
equations of motion for the fields $\H_I$ and $B_I$:

\begin{eqnarray}
  && \frac{\d \La}{\d \H_I} = 0 \quad \Longrightarrow \quad \mez
  \M^{IJ} {}^* \H_J = S^I + A^I - \r^I \label{eomh2} \\
  && \frac{\d \La}{\d B_I} = 0 \quad \Longrightarrow \quad \mm^{IJ}
  {}^* B_J + \met^{IJ} B_J = T^I - \de \r^I \label{eomb2}
\end{eqnarray}

\noi and their Hodge dual:

\begin{eqnarray}
  && \mez \M^{IJ} \H_J = {}^* \sx( S^I + A^I - \r^I \dx)
  \label{eomhs}\\  
  && \mm^{IJ} B_J + \met^{IJ} {}^* B_J = {}^* \de \r^I - {}^* T^I 
  \label{eombs} 
\end{eqnarray}

\noi equations (\ref{eomh2}) and (\ref{eomhs}) can be immediately
inverted to obtain the expression for $\H_I$ (respectively ${}^*
\H_I$) in terms of ${}^* \r^I$ ($\r^I$) while the usual linear
combination of equations (\ref{eomb2}) and (\ref{eombs}) gives the
expression for $B_I$ (${}^* B_I$) in terms of $\de \r^I$ and ${}^* \de
\r^I$: 

\begin{eqnarray}
  \H_I &=& 2 \M_{IJ} {}^* \sx( S^J + A^J - \r^J \dx) \label{hr} \\ 
  {}^* \H_I &=& 2 \M_{IJ} \sx( S^J + A^J - \r^J \dx) \label{hsrs} \\
  B_I \!&=&\! \mqt^{-1}_{IJ} \sx[ \sx( {}^* T^J - \etsm^J_K T^K \dx) -
  \sx( {}^* \de \r^J - \etsm^J_K \de \r^K \dx) \dx] \label{bdr} \\   
  {}^* B_I \!&=&\! \mqt^{-1}_{IJ} \sx[ \sx( \de \r^J + \etsm^J_K {}^*
  \de \r^K \dx) - \sx( T^J + \etsm^J_K {}^* T^K \dx) \dx]
  \label{bsdrs} 
\end{eqnarray}

\noi Notice that there are again some sign differences with equations
(\ref{brho}), (\ref{bsrho}) due to the new definition of $\mm^{IJ}$,
$\eet^{IJ}$ and $\etsm^I_J$ with $\im$ instead of $\d$. By substituting
(\ref{bdr}--\ref{hsrs}) in (\ref{lagIV}), one gets the dualized
Lagrangian:

\begin{eqnarray}
  \label{ldual}
  \La_{Dual} &=& \M_{IJ} \sx( {}^* A^I \w A^J + \M_{IJ} {}^* \r^I \w
  \r^J - 2 \M_{IJ}{}^* A^I \w \r^J \dx) + \nn \\
  && + 2 \M_{IJ} \sx( {}^* A^I + {}^* \r^I \dx) \w S^J + \M_{IJ} {}^*
  S^I \w S^J + \nn \\ 
  && - \mez \mqt^{-1}_{IJ} \de {}^* \r^I \w \de \r^J + \mez
  \mqt^{-1}_{IK} \etsm^K_J \de \r^I \w \de \r^J + \nn \\ 
  && + \mqt^{-1}_{IJ} {}^* T^I \w \de \r^J - \mqt^{-1}_{IK} \etsm^K_J
  T^I \w \de \r^J + \nn \\
  && - \mez \mqt^{-1}_{IJ} {}^* T^I \w T^J + \mez \mqt^{-1}_{IK}
  \etsm^K_J T^I \w T^J 
\end{eqnarray}

\noi It should be noticed that, in the absence of Fayet-Iliopoulos
terms ($e^I_\L = m^{I\L} = 0$), the equations of motion for $\H_I$
would have given the standard $N=2$ Lagrangian if $\de q^I = - \r^I$, 
where the minus sign is due to the opposite sign of the Lagrange
multiplier term in the standard Lagrangian, when dualizing $\na q^I$
(see reference \cite{Dall'Agata:2003yr}) with respect to that in
(\ref{lagIV}). This Lagrangian describes a system made up by $n$
massive vectors, coupled to scalars ($A^I$) and fermions ($T^I$).
Together with all the undualized terms of $N=2$ $D=4$ Lagrangian of
\cite{D'Auria:2004yi}, it describes a supergravity theory with massive
vectors, which can not be identified with a standard $N=2$ $D=4$
supergravity, due both to the presence of a number of massive vectors,
and to the fact that the vector- and hyper- multiplets are not written
in the standard way.\\ 
In order to recover a more usual Lagrangian, we make use of the
St\"uckelberg mechanism, which allows one to rewrite the mass terms as
kinetic terms for a number of scalars coupled to vectors, with a gauge
invariance in order to keep the correct number of degrees of freedom
fixed.\\ 
The St\"uckelberg mechanism works as follows: at first it has to be
noted that the kinetic term for the vectors is invariant under gauge 
transformations: 

\begin{equation}
  \d \r^I = \de \O^I \label{gaug2}
\end{equation}

\noi where $\O^I$ is a 0--form gauge parameter. On the contrary the
mass terms, in which the bare field $\r^I$ is present, are not
invariant under (\ref{gaug2}).\\ 
In order to extend this invariance to the whole Lagrangian, we can
think of the mass terms as being already gauge-fixed, as in equation 
(\ref{redg}), so that it is possible to recover the gauge invariance
introducing a new 0--form field $\phi^I$ together with a massless
vector $\A^\L$, whose gauge transformations leave $\r^I$ invariant. 
Summarizing we can write $\r^I$ as:

\begin{equation}
  - \r^I \equiv k^I_\L \A^\L + \de \phi^I \label{rtocov}
\end{equation}

\noi where now all is invariant under the following gauge
transformations:

\begin{eqnarray}
  \d \A^\L &=& \de \O^\L \label{delA}\\
  \d \phi^I &=& - k^I_\L \O^\L \label{delphi}
\end{eqnarray}

\noi since the $k^I_\L$ are constant. The r.h.s. of equation
(\ref{rtocov}) can also be interpreted as the covariant derivative of
the scalar field $\phi^I$, where the gauge fields are the vectors
$\A^\L$, and $k^I_\L$ have the role of Killing vectors.\\
After this redefinition, the Lagrangian reads:

\begin{equation}
  \label{lagnew}
  \La = \La_{scal}^{kin} + \La_{vect}^{kin} + \La_{scal}^{Pauli} +
  \La_{vect}^{Pauli} + \La_{inv}^{4f} + \La_{non\, inv}^{4f} 
\end{equation}

\noi where:

\begin{eqnarray}
  \La_{scal}^{kin} &=& \M_{IJ} \sx( {}^* A^I \w A^J + {}^* \na \phi^I
  \w \na \phi^J + 2 {}^* A^I \w \na \phi^J \dx) \label{lsk} \\ 
  \La_{scal}^{Pauli} &=& 2 \M_{IJ} {}^* A^I \w S^J + 2 \M_{IJ}{}^*
  \na \phi^I \w S^J \label{lsp} \\ 
  \La_{inv}^{4f} &=& \M_{IJ} {}^* S^I \w S^J \label{ls4} \\  
  \La_{vect}^{kin} &=& - \mez \mqt^{-1}_{IJ} k^I_\L k^J_\S F^\L \w
  {}^* F^\S + \mez \mqt^{-1}_{IK} \etsm^K_J k^I_\L k^J_\S F^\L \w F^\S 
  \label{lvk} \\
  \La_{vect}^{Pauli} &=& - \mqt^{-1}_{IJ} k^J_\L T^I \w {}^* F^\L +
  \mqt^{-1}_{IK} \etsm^K_J k^J_\L T^I \w F^\L \label{lvp} \\ 
  \La_{non\, inv}^{4f} &=& - \mez \mqt^{-1}_{IJ} T^I \w {}^* T^J +
  \mez \mqt^{-1}_{IK} \etsm^K_J T^I \w T^J \label{lv4} 
\end{eqnarray}

\noi where as usual $F^\L \equiv \de \A^\L$.\\
The kinetic scalar terms of equation (\ref{lsk}), together with the
$g_{uv} {}^* \de q^u \w \de q^v$ term of
\cite{Dall'Agata:2003yr,D'Auria:2004yi} give back the standard kinetic
term for the hyperscalars, provided we identify: 

\begin{equation}
  \label{phiq}
  \na \phi^I \equiv \na q^I.
\end{equation}

\noi The Pauli terms of equation (\ref{lsp}) can be divided into two
sets: those proportional to the ``rectangular'' vielbein $\U_{IA\a}$, 
and those proportional to the connection $\D_I{}^{\a}{}_\b$. The former
join the analogue terms proportional to $P_{uA\a}$ in
\cite{Dall'Agata:2003yr,D'Auria:2004yi},  giving rise to the standard
$N=2$ Pauli terms coupling the hyperscalars to the hyperinos.\\
The latter is necessary in order to reconstruct the standard covariant
derivative of the hyperinos. Indeed the kinetic term of the hyperinos
in the Lagrangian of \cite{D'Auria:2004yi} has, as pointed out in 
\cite{Theis:2003jj} in the ungauged case, the following symplectic
connection term:

\begin{equation}
  \i \ol \z^\a \de q^u \D_{u\a}{}^\b \z_\b + \cc \label{conndual}
\end{equation}

\noi where $\D_{u\a}{}^\b = {\hat \D}_{u\a}{}^\b - A^I_u {\hat
\D}_{I\a}{}^\b $, the hatted quantities being the quaternionic ones.\\ 
Adding to (\ref{conndual}) the terms in (\ref{lsp}) we get:

\begin{equation}
  \de q^u {\hat \D}_{u\a}{}^\b - \de q^u A^I_u {\hat \D}_{I\a}{}^\b +
  \de q^u A^I_u {\hat \D}_{I\a}{}^\b + \na q^I {\hat \D}_{I\a}{}^\b =
  \na q^{\hat u} {\hat \D}_{{\hat u}\a}{}^\b \label{syconn}
\end{equation}

\noi Where $\hat \D_{\hat u \a}{}^\b$ is the standard $N=2$ gauged
supergravity Sp$(2n_H)$ connection of the quaternionic manifold.
Finally the 4--fermion terms of (\ref{ls4}) are the same
appearing as a result of the dualization of the scalars into tensors in
the Lagrangian of reference \cite{D'Auria:2004yi}, with the opposite
sign: the total contribution of both terms is vanishing, that is they
are only rearrangements of the bilinear fermions caused by the
dualization procedure.\\ 
Let us now turn to the last three terms of (\ref{lagnew}): they have
the expected form to be interpreted as standard $N=2$ supergravity
Lagrangian terms, except for their coupling matrices. The vector
kinetic terms of (\ref{lvk}) have the correct form, provided we make the 
following redefinition of the period matrix: 

\begin{eqnarray}
  \mqt^{-1}_{IJ} k^I_\L k^J_\S &=& \sx( {\rm Im} \N' \dx)_{\L\S}
  \label{identim} \\
  \mqt^{-1}_{IK} \etsm^K_J k^I_\L k^J_\S &=& \sx( {\rm Re} \N'
  \dx)_{\L\S} \label{identre} 
\end{eqnarray}

\noi The Pauli terms in (\ref{lvp}) can be treated collectively,
indeed let us write $T^I$ and its Hodge dual as: 

\begin{eqnarray}
  T^I &=& \im_{\L\S} m^{I\L} \sx[ X^\S_a \chi^{a-} + \ol X^\S_a
  \chi^{a+} \dx] \label{tidef}\\ 
  {}^* T^I &=& \i \, \im_{\L\S} m^{I\L} \sx[ X^\S_a \chi^{a-} - \ol
  X_a^\S \chi^{a+} \dx] \label{stidef}
\end{eqnarray}

\noi where $\chi^{a\pm}$ ($\chi^{a+} = \sx( \chi^{a-} \dx)^*$) is a
fermion bilinear and $X_a^\L$ is the upper part of a symplectic vector:

\begin{equation}
  X_a^\L \equiv \sx\{ 
  \begin{array}[]{l}
    L^\L \cr \ol f^\L_\ib
  \end{array}
  \dx. , \qquad \ol X_a^\L \equiv \sx\{ 
  \begin{array}[]{l}
    \ol L^\L \cr f^\L_i
  \end{array}
  \dx.\label{sysec} 
\end{equation}

\noi gathering the common factors, they read: 

\begin{eqnarray}
  \La^{Pauli}_{vect} &=& \mqt^{-1}_{IK} F^\L k^I_\L \sx[ \sx( - \i
  \im_{\D\Pi} m^{J\Pi} \d^K_J + \etsm^K_J \im_{\D\Pi} m^{J\Pi} \dx)
  X^\D_a \chi^{a-} + \cc \dx] \nn \\ 
  && = \mqt^{-1}_{IK} F^\L k^I_\L \sx[ \sx( e^K_\D - \N_{\D\S} m^{K\S}
  \dx) X^\D_a \chi^{a-} + \cc \dx] \label{pauli} 
\end{eqnarray}

\noi so that we are forced to introduce the new symplectic vectors: 

\begin{eqnarray}
  {L'}^\L &\equiv& \sx( k^{-1} \dx)^\L_I \sx( e^I_\Pi - \N_{\Pi\S}
  m^{I\S} \dx) L^\Pi\label{newsyl} \\ 
  {\ol f'}^\L_\ib &\equiv& \sx( k^{-1} \dx)^\L_I \sx( e^I_\Pi -
  \N_{\Pi\S} m^{I\S} \dx) {\ol f}^\Pi\ib \label{newsyf} 
\end{eqnarray}

\noi in terms of which (\ref{lvp}) reads:

\begin{eqnarray}
  \La^{Pauli}_{vect} &=& F^{\L-}_{\m\n} \sx( \im' \dx)_{\L\S} \sx[ 4
  {L'}^\S \sx( \ol \psi^{A|\m} \psi^{B|\n} \e_{AB} \dx)^- - 4 \i
  {f'}^\S_\ib \sx( \ol \l^\ib_A \g^\m \psi_B^\n \e^{AB} \dx)^- +
  \dx. \nn \\ 
  && \sx. + \mez \na_i {f'}_j^\S \ol \l^{iA} \g^{\m\n} \l^{jB} \e_{AB}
  - {L'}^\S \ol \z_\a \g^{\m\n} \z_\b \IC^{\a\b} + \cc \dx]
  \label{lag4f} 
\end{eqnarray}

\noi Note that $L'$ and $\ol f'$ can be shown to be covariantly
holomorphic as following from the Special Geometry, so that the
redefinitions (\ref{newsyl}), (\ref{newsyf}) are consistent.\\ 
Let us now turn to the 4--fermion terms; it should be noticed that,
together with the terms of (\ref{lv4}), there are the non-invariant
terms already present in \cite{Andrianopoli:1996cm} in order to have a
supersymmetric theory; these terms can collectively be written (see
\cite{Ceresole:1995jg}): 

\begin{equation}
  \La^{4f} = \frac{\i}{4} \im_{\L\S} X^\L_a X^\S_b \chi^{a-} \chi^{b-}
  + \cc \label{l4f}
\end{equation}

\noi while those coming from dualization turn out to be:

\begin{eqnarray}
  \La^{4f}_{non\, inv} &=& - \mez \mqt_{IK}^{-1} \sx(\i \d^K_J
  \im_{\D\Pi} m^{J\D} - \etsm^K_J \im_{\D\Pi} m^{J\D} \dx) \im_{\L\S}
  m^{I\L} \cdot \nn \\ 
  && \qquad \cdot \sx[ X^\S_a \chi^{a-} + \ol X^\S_a \chi^{a+} \dx]
  \sx[ X^\Pi_b \chi^{b-} + \ol X^\Pi_b \chi^{b+} \dx] \label{4f}  
\end{eqnarray}

\noi summing (\ref{l4f}) and (\ref{4f}), after some algebraic
manipulation we get the following expression:

\begin{equation}
  \La^{4f} = \frac{\i}{4} {\im'}_{\L\S} {X'}^\L_a {X'}^\S_b \chi^{a-}
  \chi^{b-} + \cc \label{4ftot}
\end{equation}

\noi which is the expected 4--fermions Lagrangian in terms of the new
symplectic sections and period matrix.

\section{Symplectic rotation}
\label{syrot}

The final expression for the dual Lagrangian is therefore that of a 
standard $N=2$ gauged supergravity \cite{Andrianopoli:1996cm},
provided the formulae (\ref{identim}--\ref{identre}) and
(\ref{newsyl}--\ref{newsyf}) define respectively a period matrix and
the upper components of a symplectic section.\\
Let us consider the transformation which, acting on the charges
vector, rotates it into a vector of purely electric charges, that is:

\begin{equation}
  \label{rot}
  \begin{pmatrix}{m^{I\L} \cr e^I_\L }\end{pmatrix} \to
  \begin{pmatrix}{0 \cr k^I_\L }\end{pmatrix}.
\end{equation}

\noi This transformation is performed by means of the matrix $S$:

\begin{equation}
  \label{defs}
  S = \begin{pmatrix}{\sx( k^{-1} \dx)^\S_I e^I_\L & - \sx( k^{-1}
  \dx)^\S_I m^{I\L} \cr 0 & k^I_\S \sx( e^{-1} \dx)^\L_I }\end{pmatrix}
\end{equation}

\noi where we have supposed that the matrix $k^I_\L$ is invertible, as
well as $m^{I\L}$ and $e^I_\L$. The effect of $S$ on the period matrix
turns out to be:

\begin{equation}
  {\N'}_{\L\S} = k^J_\L \sx( e^{-1} \dx)^\Pi_J \N_{\Pi\D} \sx[ \sx(
  k^{-1} \dx)^\S_I e^I_\D - \sx( k^{-1} \dx)^\S_I m^{I\G} \N_{\G\D}
  \dx]^{-1} \label{sn} 
\end{equation}

\noi According to equation (\ref{sn}) the real and the imaginary part
of the rotated period matrix are:

\begin{eqnarray}
  \sx( {\rm Im} \N' \dx)_{\L\S} &=& \mqt^{-1}_{IJ} k^I_\L k^J_\S 
  \label{impr} \\
  \sx( {\rm Re} \N' \dx)_{\L\S} &=& \mqt^{-1}_{IK} \etsm^K_J k^I_\L
  k^J_\S \label{repr}  
\end{eqnarray}

\noi This same transformation will act also on the symplectic sections
according to:  

\begin{equation}
  \begin{pmatrix}{{L'}^\L \cr {M'}_\L}\end{pmatrix} \equiv
  \begin{pmatrix}{\sx( k^{-1} \dx)^\S_I e^I_\L & - \sx( k^{-1}
      \dx)^\S_I m^{I\L} \cr 0 & k^I_\S \sx( e^{-1} \dx)^\L_I
    }\end{pmatrix} \begin{pmatrix}{L^\L \cr M_\L}\end{pmatrix}
    = \begin{pmatrix}{ \sx( k^{-1} \dx)^\L_I \sx( e^I_\L - \N_{\L\S}
    m^{I\S} \dx) L^\L \cr k^I_\S \sx( e^{-1} \dx)^\L_I
    M_\L}\end{pmatrix} \label{synew}
\end{equation}

\noi and similarly for the symplectic section
$\begin{pmatrix}{{f'}^\L_i \cr {h'}_{\L |i}}\end{pmatrix}$. Therefore, 
modulo a total derivative term $(e^{-1})^I_\L (m^{-1})_{I\S} F^\L
F^\S$ which does not modify the equations of motion, the
transformation $S$ acting on the symplectic sections and period matrix
reproduces the expressions of equations (\ref{identim}--\ref{identre})
and (\ref{newsyl}--\ref{newsyf}); this implies that the effect of the
introduction of the Fayet-Iliopoulos terms $m^{I\L}$ and $e^I_\L$ in
the Lagrangian coupled to tensor multiplets corresponds, after the
dualizations of the massive tensors to massive vectors, and the
subsequent reinterpretation of these latter as massless vectors and
scalars via the St\"uckelberg mechanism, to a purely electric gauging
of the undualized standard Lagrangian, together with a rotation of all
the symplectic sections and the period matrix with the matrix $S$.

\section{Conclusions}
\label{conc}

In this paper we have shown that, in an even number $D$ of dimensions,
an arbitrary number $n$ of $D/2$ massive tensor fields, with masses
given by both electric and magnetic couplings, are dual to $n$ massive
$(D/2-1)$ tensors, whose masses are only of the electric type and are
given by relation (\ref{massq}). The formalism developed to show this
duality has been applied to the Lagrangian of the $N=2$ $D=4$ supergravity
coupled to $n$ tensor and $n$ vector multiplets, showing that it is
dual to a standard $N=2$ $D=4$ gauged supergravity where only electric
charges are present.\\
The Lagrangian obtained in section \ref{sec:stsugra} is related to the
standard formulation of $N=2$ $D=4$ gauged supergravity
\cite{Andrianopoli:1996cm} by a symplectic rotation, defined in
(\ref{defs}) acting on all the Special Geometry quantities. Since
however a symplectic rotation is  not a symmetry of the theory, unless
in the generical symplectic matrix $T= \begin{pmatrix}{A & B \cr C &
D}\end{pmatrix}$ the blocks $B$ and $C$ are set to 0
\cite{Andrianopoli:1996cm}, the two theories related by the symplectic
rotation (\ref{defs}) are not the same theory, as is evident by
looking at their $N=1$ truncations which give rise to different theories
\cite{wip}. 

\acknowledgments{I would like to thank R. D'Auria and M. Trigiante for
useful discussions and suggestions. Work supported in part by the
European Community's Human Potential Program under contract
MRTN-CT-2004-005104 'Constituents, fundamental forces and symmetries
of the universe', in which L.S. is associated to Torino University.}


\end{document}